\documentclass{emulateapj}	
\usepackage{times}
\usepackage{graphicx}

\slugcomment{To Appear in ApJL }

\def\lsim{\mathrel{\rlap{\lower4pt\hbox{\hskip1pt$\sim$}}
    \raise1pt\hbox{$<$}}}                
\def\gsim{\mathrel{\rlap{\lower4pt\hbox{\hskip1pt$\sim$}}
    \raise1pt\hbox{$>$}}}                

\shorttitle{PSR J1311$-$3430}
\shortauthors{Romani et al.}

\begin{document}

\title{PSR J1311$-$3430: A Heavyweight Neutron Star with a Flyweight Helium Companion}

\author{Roger W. Romani\altaffilmark{1}, Alexei V. Filippenko\altaffilmark{2},
Jeffery M. Silverman\altaffilmark{2}, S. Bradley Cenko\altaffilmark{2}, 
Jochen Greiner\altaffilmark{3}, Arne Rau\altaffilmark{3}, Jonathan Elliott\altaffilmark{3},
and Holger J. Pletsch\altaffilmark{4}}
\altaffiltext{1}{Department of Physics, Stanford University, Stanford, CA 94305, USA; rwr@astro.stanford.edu}
\altaffiltext{2}{Department of Astronomy, University of California, Berkeley, CA 94720-3411, USA}
\altaffiltext{3}{Max-Planck-Institut f\"ur Extraterrestrische Physik, D-85748 Garching, Germany}
\altaffiltext{4}{Max-Planck-Institut f\"ur Gravitationsphysik (Albert-Einstein-Institut), 
D-30167 Hannover, Germany; Institut f\"ur Gravitationsphysik, 
Leibniz Universit\"at Hannover, D-30167 Hannover, Germany}

\begin{abstract}

We have obtained initial spectroscopic observations and additional
photometry of the newly discovered $P_b=94$\,min $\gamma$-ray black-widow pulsar
PSR J1311$-$3430. The Keck spectra show a He-dominated, nearly H-free photosphere
and a large radial-velocity amplitude of $609.5\pm7.5$\,km\,s$^{-1}$. 
Simultaneous seven-color GROND photometry further probes the heating
of this companion, and shows the presence of a flaring infrared excess.
We have modeled the quiescent light curve, constraining the orbital inclination and masses.
Simple heated light-curve fits give $M_{\rm NS}=2.7\,{\rm M}_\odot$, but show systematic light-curve
differences. Adding extra components allows a larger mass range to be fit,
but all viable solutions have $M_{\rm NS}>2.1\,{\rm M}_\odot$. If confirmed, such a large $M_{\rm NS}$ 
substantially constrains the equation of state of matter at supernuclear densities.
\end{abstract}


\section{Introduction}

	The bright $\gamma$-ray source 2FGL J1311.7$-$3429 has
been known, and unidentified, since the early EGRET mission
(Fichtel et al. 1994).  After \citet[][hereafter R12]{r12} discovered an optical counterpart with
a $P_b=5626.0$\,s (93.8\,min) orbital period and evidence for very strong pulsar heating,
\citet{pet12} managed to discover 2.5\,ms Doppler-modulated pulsations 
in the {\it Fermi} Large Area Telescope \citep[LAT,][]{LAT09} $\gamma$-ray photons. Thus
with orbital constraints, $\gamma$-only millisecond pulsars can be 
detected by the LAT, although PSR J1311$-$3430 was later found to also be
intermittently visible in the radio (Ray et al. 2012, in prep.). Additional optical
and X-ray observations supporting the orbital variability found above
are described by \citet{ket12}.


The $\gamma$-ray pulsar ephemeris gives an orbit of $a_{\rm NS}\,{\rm sin}\,i = 0.010581$\,lt-s	
for a minimum companion mass of $\sim 8 \times 10^{-3}\,{\rm M}_\odot$.
This measurement alone does not significantly constrain the neutron 
star mass, which is of particular interest, since \citet[][hereafter vKBK]{vKBK11}
found evidence that the original, $P_b=9.2$\,hr ``black widow'' (BW) binary 
PSR B1957+20 may host a remarkably heavy neutron star 
with $M_{\rm NS}=2.4\pm 0.12\,{\rm M}_\odot$.  If BW pulsars as a class are massive, 
PSR J1131$-$3430, with its ultra-short orbital period (the smallest
of any rotation-powered pulsar), provides interesting opportunities for
a mass determination.

	We have been able to obtain exploratory spectroscopy and additional
photometry that constrain properties of the PSR J1311$-$3430 system.
These data show that the companion is a bloated, Roche-lobe filling
substellar object whose photosphere is He dominated, with no detected H.
While the neutron star mass estimate is quite model dependent, the data 
already imply that it is quite large.  We discuss here the observations, the
implications, and the prospects for further refinement of the system parameters.

\section{Keck LRIS Spectroscopy}

	Six consecutive 300\,s exposures of J1311$-$3430 were obtained with the Low Resolution
Imaging Spectrometer (LRIS; Oke et al. 1995) at the Keck-I 10\,m telescope on 2012 May 17 (UT dates are used herein; MJD 56064).
We used the $1^{\prime\prime}$ long slit at the parallactic angle (Filippenko 1982) and the 5600\,\AA\
dichroic splitter. In the blue camera the 600/4000 grism provided coverage to
$<3400$\,\AA\ at $\sim 4$\,\AA\ resolution; the red camera employed the
400/8500 grating for $\sim 7$\,\AA\ resolution. The seeing was good 
($\sim 0.8^{\prime\prime}$) and conditions were nearly photometric. Standard processing and
optimal extraction were applied; fluxes were calibrated against the red
and blue standards BD+26$^\circ$2606 and BD+28$^\circ$4211, respectively.

	These observations covered the phase of maximum light ($0.7<\phi_B<1.1$), so the 
spectrum is very blue (Figure 1). Although the temperature inferred from
the color is $\gsim$10,000\,K (see \S3), no Balmer lines are visible, implying 
an unusual surface composition. This is in contrast to the BW
companions of PSR B1957+20 \citep[vKBK;][]{arc92} and 2FG J2339-0533
\citep{rs11}, where relatively normal companion spectra are seen, with spectral
classes appropriate
to their (cooler) surface temperatures. After shifting the individual
spectra to the rest frame using the sinusoidal approximate Doppler curve
we obtain a phase-averaged spectrum dominated by narrow neutral He lines.
We infer that this is the H-stripped remnant of an evolved stellar core.

	The lower panel of Figure 1 shows the blue region of the normalized
spectrum compared with a He-dominated model from the atmosphere grid of 
\citet{jwp01}, smoothed to the resolution of the Keck data. The 
correspondence is excellent, with the exception that the Balmer lines,
well visible in the model despite the $10^{-4}$ H number abundance, are undetectable
in PSR J1311$-$3430. This implies a maximum H abundance $n({\rm H}) <10^{-5}$. 

        We have used the model grid to estimate the effective temperature
and gravity of this phase-averaged spectrum. The fit minima give 
$T_{\rm eff} =12,000\pm1000$\,K and log\,$g=4.6\pm0.2$. Measurements of the
equivalent width (EW) of individual lines
are consistent with these conclusions; for example, the ratio ${\rm EW}_{4471}/{\rm EW}_{4481} = 1.3\pm0.2$
and the absence of He~II support the $T_{\rm eff}$ estimate, while the 
ratio ${\rm EW}_{4471}/{\rm EW}_{3819} = 1.97\pm0.2$ accords with the log\,$g$ value.
A few species, such as Si $\lambda$4552 and Ni $\lambda$3995, suggest some flux at
lower values of log\,$g$. This is not very surprising, since the companion is likely
near-Roche-lobe filling with appreciable surface-gravity variation. Higher
signal-to-noise ratio (S/N)
spectra and synthetic composite spectral models will be needed to
study the atmosphere's surface variation.

\begin{figure}[t!!]
\vskip 9.3truecm
\includegraphics{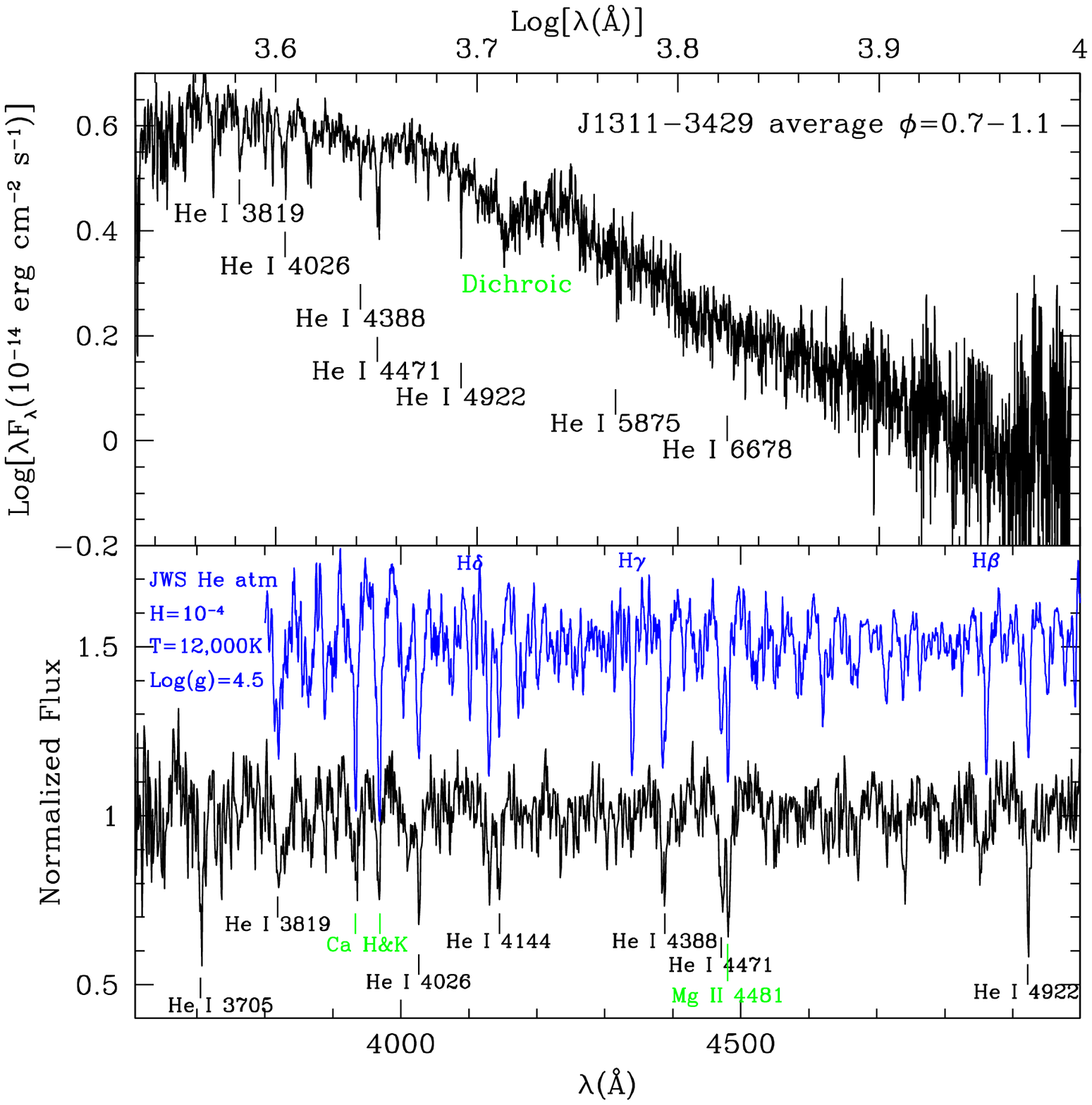}
\begin{center}
\caption{\label{LRISspec} 
Combined, phased Keck LRIS spectrum. Upper Panel: Full spectrum. Lower panel:
Spectrum showing He~I line domination and the good match to the
\citet{jwp01} model atmosphere (blue). 
}
\end{center}
\vskip -0.7truecm
\end{figure}

	The highly unusual He-dominated, low log\,$g$ spectrum compromised attempts to
find suitable cross-correlation templates. Accordingly, we adopt the best-fit atmosphere
model to measure the radial velocities, using the RVSAO package in IRAF. 
All observed comparison stars were poor spectral
matches and gave significantly lower correlation coefficients.  
However, all templates gave very similar radial velocities.


	With the phase fixed by the pulsar-determined orbital
 ephemeris, we can fit a simple sinusoid to
these velocities, obtaining amplitude $K_{\rm obs}=609.5 \pm 7.5\, {\rm km\,s^{-1}}$ and
average radial velocity $\Gamma= 62.5 \pm 4.5 \, {\rm km\,s^{-1}}$, with $\chi_\nu = 1.3$. At face value
this gives a mass function
$$
f_{\rm 2, obs} = {{K_{\rm obs}^3P_b} \over {2\pi G}}= { {(M_{\rm NS} {\rm sin}\, i)^3} \over {M_{\rm total}^2}} = 1.54 \pm 0.06\,{\rm M}_\odot
$$
and a very large mass ratio 
$q_{\rm obs} = (K_{\rm obs}/K_{\rm NS})^3 = 172.1\pm 2.1$. However, the
observed spectrum tracks the center of light (CoL) from the heated face 
of the companion, which has a smaller radial-velocity amplitude than the 
companion center of mass (CoM). For PSR B1957+20, vKBK estimate the correction factor
as $K_{\rm cor} = 1.09$ (i.e., $K_{\rm CoM}=K_{\rm cor}K_{\rm obs}$). 
If we adopt this value of $K_{\rm cor}$ and the inclination constraint $i < 85^\circ$ from 
the lack of X-ray eclipses \citep[R12;][]{ket12}, we derive $M_{\rm NS} > 2.03\,{\rm M}_\odot$. 
This is an interestingly high mass for equation-of-state (EoS) constraints.

	However, $K_{\rm cor}$ depends on the heating across the surface. Also,
\citet{rs11} find that for the similar BW-type binary J2339$-$0533, the heating
introduces a significant non-sinusoidal component to the CoL curve, and it should be fit.
Any further restrictions on the inclination $i$ greatly narrow the pulsar
mass range. These factors can be probed with light-curve modeling, so we have
sought to improve the photometry to allow better constraints on the system masses.

\section{GROND Photometry}

	R12 measured the companion light curve from WIYN, SOAR, and archival
VLT 3-band ($g^\prime r^\prime i^\prime$ or $BVI$) photometry. However, the colors were not simultaneous;
the largest dataset (SOAR Optical Imager $g^\prime r^\prime i^\prime$) covered each filter
on a separate night. These data showed significant epoch-to-epoch variability 
and large (up to 4 mag) flares, so precise measurements of the instantaneous
colors were not available. Accordingly, we targeted J1311$-$3430 with the GROND
system on the MPI/ESO 2.2\,m telescope on La Silla \citep{GROND}, obtaining
simultaneous $g^\prime r^\prime i^\prime z^\prime JHK$ images starting 2012 July 09.
The observations lasted 2.5\,hr ($\sim 1.5$ orbits) from MJD 56117.956 to
56118.059, covering two maxima (pulsar inferior conjunction). Twelve observation
blocks were made, each with $4\times 115$\,s dithered optical pointings and 
$8\times 60$\,s near-IR exposures. The seeing was $\sim 1^{\prime\prime}$ 
and the airmass increased slowly. Near maximum light, individual
$g^\prime r^\prime i^\prime$ exposures had adequate S/N, allowing a temporal resolution
of $\sim 195$\,s, while at minimum and for all $z^\prime$ images 4 dithered
exposures were combined to achieve adequate S/N. 
To ensure the best possible calibration of the optical bands, the target was revisited
on 2012 August 10 with SDSS field exposures bracketing the J1311$-$3430 observations,
allowing an accurate calibration of the field and target magnitudes; the
calibration systematics are small compared to the photometric errors, and are
included in the error flags of Figure 2 ($\sigma_{g^\prime} \approx 0.02$,
$\sigma_{r^\prime} \approx 0.03$, $\sigma_{i^\prime} \approx 0.05$ mag at maximum). 

	J1311$-$3430 had relatively low near-IR S/N and was detected at
high significance only near maximum light; we thus made photometric 
measurements on an image stacked from frames covering phase $\pm 0.17$.
The near-IR data were calibrated using 2MASS stars in the field, and 
transformed to AB magnitudes for spectral energy distribution (SED) fitting together with the optical data.
These peak magnitudes were $J = 20.6 \pm 0.1$, $H=21.1 \pm 0.3$, and $K = 20.9\pm 0.3$.
All magnitudes were corrected for the filter-specific dust-map-estimated 
Galactic extinction ($A_V=0.173$ mag), using the \citet{sf11} calibration.

Figure 2 summarizes the GROND photometry. The left panel shows the
$g^\prime r^\prime i^\prime z^\prime$ light curves. The lower portion displays
the $g^\prime-i^\prime$ color, blue at maximum and redder at minimum. The upper
right shows the optical through near-IR SED at maximum light ($-0.1<\phi_B < 0.1$)
and the optical SED near quadrature ($\phi_B=0.2$--0.3, 0.7--0.8). 
We see that at maximum the colors are comparable to those of a B7--B8 star
in the blue, but there is a large near-IR excess. Although the pulsar
heating makes the star multi-temperature, this very large excess suggests
a large emitting area at low temperature. One source larger than the
companion photosphere is the evaporative wind,
where pulsar power may be reprocessed into the optical-IR.  The optical light curves show
appreciable fluctuations on short timescales. The lower-right panel
summarizes this variability with the $g^\prime-i^\prime$ color plotted against
the $g^\prime$ magnitude. Again the star is redder at fainter magnitudes.
Interestingly, if we flag points that are brighter than the ``quiescent''
light-curve flux (see below), most of these points shift (green lines) to larger
$g^\prime-i^\prime$ (redder) relative to the expected quiescent color.  
An extreme example is the very large $i^\prime$ flare recorded by R12.
We conclude that the principal light-curve fluctuations have a red SED.
This implicates modulation in reprocessed pulsar emission by the
variable wind off the companion.
An alternative fluctuation site, stellar flares on the tidally locked companion,
would in contrast be very hot (blue).

\begin{figure}[t!!]
\vskip 8.9truecm
\includegraphics{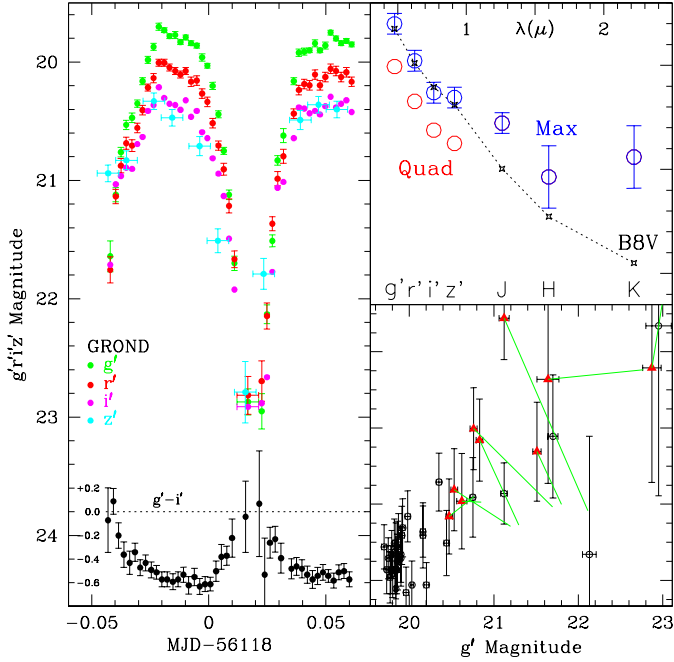}
\begin{center}
\caption{\label{GRONDres} 
GROND photometry. Left: $g^\prime r^\prime i^\prime z^\prime$ light curves, showing reddening
during minimum (pulsar superior conjunction). Upper right: The companion SED at
maximum (phases $-0.1$ to 0.1, blue points) and quadrature (phases 0.2--0.3 and 0.7--0.8, red points). 
Comparison with 
a B8 star shows a large IR excess. Bottom right: J1311$-$3430 color variations. The redder 
spectrum at minimum is visible. Green lines and red triangles mark when an epoch
has $\Delta g^\prime > 0.2$ mag over the quiescent magnitude. Such flares show substantial
increases in $g^\prime - i^\prime$.
}
\end{center}
\vskip -0.5truecm
\end{figure}
\section{ELC Modeling and Parameter Estimation}

	The GROND data provide an instantaneous color reference for 
the longer, higher precision, but non-simultaneous SOAR light curves
from MJD 56008.2-56010.4. 
Just as the GROND data exhibit red flares above the ``quiescent'' flux,
comparing SOAR $g^\prime r^\prime$ photometry from adjacent orbits reveals several
$\sim 20$\,min flaring periods.  In Figure 3 we plot the phased SOAR/GROND photometric
points. During the first period all $g^\prime r^\prime$ measurements
and errors are shown --- ``flaring'' points (at $\phi_B=0.15$--0.35 and 0.45--0.8) 
are plotted as crosses.
During the second period we display only the quiescent points, but include the
GROND $i^\prime z^\prime$ quiescent data. All SOAR $i^\prime$ were affected
by strong flaring.

\begin{figure}[b!!]
\vskip 9.3truecm
\includegraphics{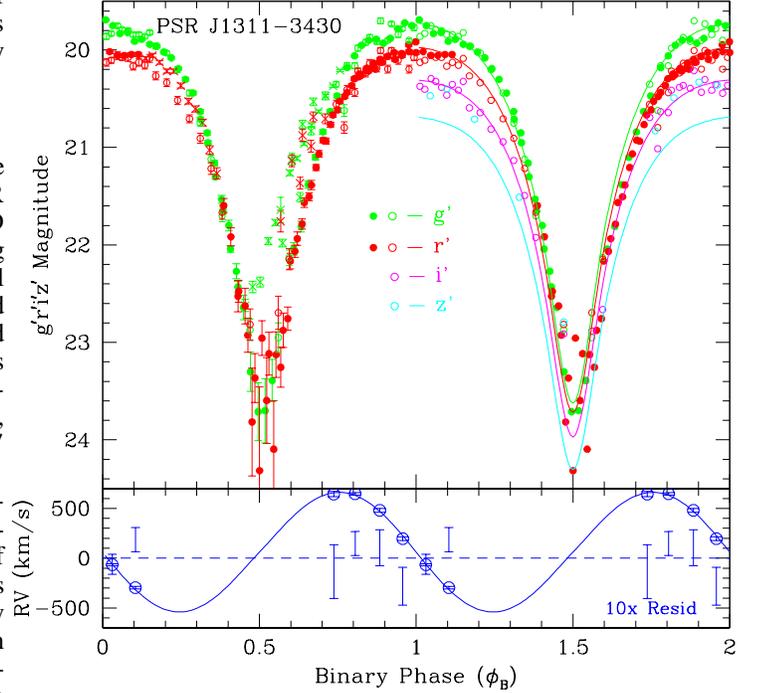}
\begin{center}
\caption{\label{LCfluxs} Top: SOAR (filled points) and GROND (open points) 
photometry; $g^\prime r^\prime$ ``flare'' measurements are flagged as crosses.
Points during the second period exclude flare epochs. Curves are from the 
``L1 Spot'' ELC fit.
Bottom: Keck radial-velocity measurements, aligned with the light curve.
Residuals are shown amplified by a factor of 10. The curve is from the same ``L1 Spot''
fit.
}
\end{center}
\vskip -0.7truecm
\end{figure}

	To constrain the system properties we have fit these ``quiescent'' 
light curves along with the radial-velocity measurements, using the ``Eclipsing Light Curve'' 
(ELC) code of 
\citet{oh00}. Typically, the code computes filter colors from atmospheres of the
``NextGen'' library \citep{hau99}, but we found that these did not match the data well, plausibly
due to the He domination of the photosphere. Instead, we fit using blackbody 
emissivities which gave reasonable light curves and colors. All fits used the
$\gamma$-ray-determined phasing and pulsar projected semimajor axis.

There are five basic fit parameters: the system mass ratio $q=M_{\rm NS}/M_c$,
the pulsar irradiation modeled here
as an isotropic flux $L_X$, the companion's underlying 
temperature $T_1$ and Roche-lobe fill factor $f_1$, and the orbital inclination $i$. $L_X$ is quite well
constrained by the $g^\prime - r^\prime$ and $r^\prime - i^\prime$ colors near maximum
light, giving log\,$(L_X)=35.3$. This is in accord with the spectroscopically
estimated $T_{\rm eff}=$ 12,000\,K near maximum.  We also find that the secondary is very close
to Roche-lobe filling in all acceptable fits, so we set $f_1=0.99$. We therefore adjusted
the three remaining parameters to fit the light curves and spectral points. The very small
photometric errors near maximum, coupled with stochastic variability, meant that
even the best fits had formal $\chi^2/dof \approx 9$, so the $T_1$ and $i$
error estimates come from the range giving a $\Delta \chi^2$ increase of $(\chi^2/dof)_{\rm min}$ 
around the fit minimum. For $q$ (and $M_{\rm NS}$) the value (for a given fit $i$) is
controlled by the radial-velocity measurements. These had $(\chi_v^2/dof)_{\rm min} \approx 1.5$,
so the error range was estimated from this smaller $\chi_v^2$ increase.

	For both PSR B1957+20 \citep{ret07} and J2339$-$0533 \citep{rs11} 
the photometric data are limited, and simple models with a pulsar-heated hemisphere 
gave adequate light-curve fits.  In contrast, the flat light-curve maxima of J1311$-$3430 
produce relatively poor fits to such models (R12). The best fit (Table 1) has a
$60^\circ$ inclination, resulting in a rather large pulsar mass, but the peak is too narrow,
the minimum is too bright by $\sim 1$\,mag, and the predicted $K_{\rm obs}$ is $2.2\sigma$ 
below the measured value. 

Adding equatorial hot spots to the companion surface bracketing the 
L1 point (30$^\circ$ radius at $35^\circ$ and $-65^\circ$) can broaden 
the light-curve peak. This mimics equatorially concentrated heating, which may be plausibly
invoked, as there is strong equatorial concentration in pulsar wind nebula (PWN) flows,
giving rise to PWN tori. Alternatively, these asymmetric components may represent
reprocessed pulsar flux in a companion wind outflow.
With these added components the model is a good match to the
$g^\prime r^\prime i^\prime$ light curves. The heated equator allows slightly 
lower log\,$(L_X)=35.0$, but the $K_{\rm cor}=1.06$ is large and the best-fit inclination is
small, giving a very (possibly unphysically) large mass.

	If, in contrast, we {\it cool} the region near L1,
by applying a $40^\circ$ spot with 1/2 the local $T$, we also 
flatten the maxima. In this case, the CoL moves toward the CoM,
resulting in a small $K_{\rm cor}=1.04$ and larger $i$. These two
effects allow a neutron star mass of $2.15\,{\rm M}_\odot$. The fit
quality is good, albeit with higher $\chi^2$ than the equatorial spot model. This
scenario gives the smallest $M_{\rm NS}$ of any viable fit. Such decreased
L1 flux might plausibly arise from large limb and gravity darkening effects
in the He-dominated atmosphere. We thus attempted to flatten the light curve by 
invoking extreme gravity darkening coefficients, as have been claimed 
for some semidetached binaries \citep{NK92}. This allowed intermediate inclination
and masses; for example, for $\tau_{\rm gr} = 0.5$, we find $i=63^\circ$ and
$M_{\rm NS}=2.30\,{\rm M}_\odot$. For all cases the fit models
underpredict the $z^\prime$ flux. This is likely related to the extra IR
component seen in Figure 2.

\begin{deluxetable}{llll}
\tablecaption{\label{Params} ELC Fit Parameters}
\tablehead{
\colhead{Parameter} & \colhead{Basic LC}& \colhead{L1 Cold Spot} & \colhead{Eq Hot Spots}
}
\startdata
$i [^\circ]$      &  60.4$\pm0.4$ & 67.3$\pm0.3$  & 57.9$\pm0.3$\cr
$T_1$[K]          &  3440$\pm50$  & $<$2000       & $<1600$     \cr
$q$               & 179.7$\pm3.9$ & 177.1$\pm3.2$ & 180.2$\pm3.3$\cr
$M_{\rm NS}[{\rm M}_\odot]$ &  2.68$\pm0.14$& 2.15$\pm0.11$ & 2.92$\pm0.16$\cr
$K_{\rm cor}$         &  1.06& 1.04                   & 1.06
\enddata
\end{deluxetable}

\section{Discussion and Conclusions}

	It is clear that additional physics is needed to fully model the light curve.
Certainly, specific intensities from an appropriate He model grid would be useful.
However, a good fit may also require some asymmetry in the heating, or re-processing 
of pulsar flux to produce optical/IR emission from the companion wind. We conclude
that, at present, model assumption systematics dominate the 
statistical fit errors and preclude accurate mass determination. Nevertheless, the large
mass function of PSR J1311$-$3430 virtually guarantees a large $M_{\rm NS}$.
In this it joins B1957+20, supporting the suggestion of vKBK that BW
pulsars as a class may have high masses.  The challenge is that secondary spectroscopy-derived
estimates for these BW masses inevitably have substantial systematic
uncertainty. In this sense, the mass estimates cannot match the ``gold standard''
timing-based $M_{\rm NS} =1.97 \pm 0.04\,{\rm M}_\odot$ for PSR J1614$-$2230 \citep{det10}.
However, since EoS constraints tighten rapidly as the minimum required
$M_{\rm NS}$ exceeds $2\,{\rm M}_\odot$, it is worth considering the prospects for
refining the J1311$-$3430 mass estimate.

	Figure 4 shows the present situation in the mass-mass plane. Compared to
PSR B1957+20, J1311$-$3430 has a larger observed mass function ($1.54\,{\rm M}_\odot$ vs. 
$1.34\,{\rm M}_\odot$). However, our models give substantially smaller $K_{\rm cor}$ than 
assumed for PSR B1957+20, so in this sense our mass fits are conservative. The
remaining systematic uncertainties in the light-curve shape allow
inclinations $57^\circ < i < 67^\circ$. This large range dominates the systematic 
mass uncertainty. The best hope for eliminating otherwise plausible models
and reducing the systematics lies with additional spectroscopy. Full-orbit 
coverage will, at minimum, substantially decrease the uncertainty in $q$. We
also expect to detect the non-sinusoidal radial-velocity component, and
with sufficient spectral resolution, the variations in absorption-line profile
expected as the visible heated surface varies. Additional simultaneous multicolor
photometry from many epochs could further isolate the photospheric light curve from
a variable wind component. Deep multicolor images can constrain the residual
illumination at pulsar superior conjunction, with additional $i$ constraints.
Of course, improved models are needed to fully exploit such data.

\begin{figure}[t!!]
\vskip 7.5truecm
\includegraphics{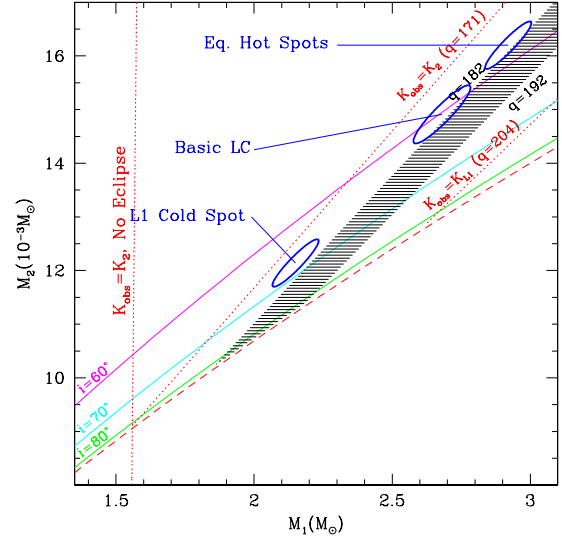}
\begin{center}
\caption{\label{LCfluxs} 
Mass constraints on the PSR J1311$-$3430 system.  The pulsar mass
function gives the solid diagonal lines for various inclinations $i$.
The dotted vertical line gives the minimum from the companion mass
function. The $K_{\rm corr}$ increase of the observed radial velocity
is critical: we show the physically allowed range from the Keck data
(dotted diagonal red lines, labeled by the appropriate value of $q$
for J1311$-$3430), along with the range for PSR B1957+20 preferred by
vKBK ($K_{\rm corr} = 1.06$--1.13; shaded region).  Fit regions for
the three models in Table 1 are indicated with blue ellipses.
}
\end{center}
\vskip -0.7truecm
\end{figure}

	Our data also illuminate several other aspects of this unique system.
The IR-dominated variability likely probes the extremely dense wind flux. Correlation
with periods of radio-pulse visibility could be very revealing.  The ultra-low-mass 
He companion points to unusual evolution, with
substantial mass transfer and irradiative stripping of the core of an evolved secondary.
\citet{BDH12} describe such a scenario, giving parameters remarkably similar to those
of J1311$-$3430. This model's slow mass transfer should allow substantial neutron star mass 
growth. Interestingly, for the large neutron star
masses indicated here, the EoS must be very ``stiff.'' This
means that for $M_{\rm NS} \approx 2.5\,{\rm M}_\odot$, we can have a moment of inertia
as large as $\sim 4 \times 10^{45}\, {\rm g\,cm^2}$ \citep{lp07}. This increases the 
spindown power of J1311$-$3430 from ${\dot E} =I \Omega {\dot \Omega} = 5 \times 10^{34}\, {\rm erg\,s^{-1}}$
(for $I_{45}=1$), accommodating the large heating flux inferred from 
the peak colors and spectroscopic $T_{\rm eff}$, even without equatorial concentration.

	In summary, PSR J1311$-$3430 presents a number of unique properties that
make it a potential Rosetta Stone for irradiation-driven evolution in tight 
binaries. If we can reduce the allowed model space, it seems likely to provide
some of the most important constraints on the EoS of dense matter.
And, even if PSR J1311$-$3430 or PSR B1957+20 do not provide the final word on 
the EoS, it is worth remembering that {\it Fermi} has proven to be an excellent finder 
of short-$P_b$ BW systems, including detection via $\gamma$-ray-blind searches. 
Future studies of this extreme population may well provide a bullet-proof case
for a stiff EoS.

\medskip
\medskip

We thank Jerry Orocz for helpful discussions on ELC model fitting,
Simon Jeffery for alerting us to the He model atmospheres, and Vlad
Sudilovsky for obtaining the GROND calibration observations.  This
work was partially supported by NASA grant NNX11AO44G. Part of the
funding for GROND (both hardware and personnel) was generously
allocated from the Leibniz-Prize to Prof. G. Hasinger (DFG grant HA
1850/28-1).  A.V.F.'s group at U.C. Berkeley is supported by Gary \&
Cynthia Bengier, the Richard \& Rhoda Goldman Fund, the Christopher
R. Redlich Fund, the TABASGO Foundation, and NSF grants AST-0908886
and AST-1211916.  Some of the data presented herein were obtained at
the W. M. Keck Observatory, which is operated as a scientific
partnership among the California Institute of Technology, the
University of California, and NASA; the Observatory was made possible
by the generous financial support of the W. M. Keck Foundation.


\end{document}